\documentclass[%
 reprint,
 amsmath,amssymb,
 aps,
]{revtex4-2}

\usepackage{graphicx}% Include figure files
\usepackage{dcolumn}% Align table columns on decimal point
\usepackage{bm}% bold math
\usepackage{color}

\def\s{{\sigma}}
\def\e{{\varepsilon}}

\def\0{{ {\bm 0} }}

\def\w{{\omega}}
\def\a{{\alpha}}

\allowdisplaybreaks[4]

\makeindex

\begin{document}

\title{
Emergence of edge ferromagnetism  and ferromagnetic fluctuations driven by the Andreev bound state in $d$-wave superconductors
}

\author{Shun Matsubara}
\author{Hiroshi Kontani}
\affiliation{Department of Physics, Nagoya University, 
Nagoya 464-8602, Japan}

\date{\today}

\begin{abstract}
As the surface Andreev bound state (ABS) forms at the open ($1,1$) edge of a $d_{x^2-y^2}$-wave superconductor,
the local density of states (LDOS) increases.
Therefore, a strong electron correlation and drastic phenomena may occur.
However, a theoretical study on the effects of the ABS
on the electron correlation has not been performed yet.
To understand these effects,
we study large cluster Hubbard model with an open ($1,1$) edge in the presence of a bulk $d$-wave gap.
We calculate the site-dependent spin susceptibility
by performing  random-phase-approximation (RPA)
and modified fluctuation-exchange (FLEX) approximation in the real space.
We find that near the ($1,1$) edge, drastic ferromagnetic (FM) fluctuations occur owing to the ABS.
In addition, as the temperature decreases,
the system rapidly approaches
a magnetic-order phase
slightly below the transition temperature of the bulk $d$-wave superconductivity (SC).
In this case, the FM fluctuations are expected to induce
interesting phenomena such as edge-induced triplet SC
and quantum critical phenomena.
\end{abstract}


%\pacs{
%74.72.Kf, 74.20.-z, 74.40.Kb, 75.25.Dk
%}

\maketitle

%%%%%%%%%%%%%%%%%%%%%%%%%%%%%%%
\section{Introduction}
In bulk cuprate superconductors, strong antiferromagnetic (AFM) fluctuations cause interesting phenomena.
For example, $d$-wave SC
\cite{Bickers,Monthoux_FLEX,Koikegami_FLEX,Takimoto_FLEX,Dahm_FLEX,Manske_FLEX}
and non-Fermi liquid phenomena in the normal state
\cite{Moriya,Moriya-2,Pines,Kontani-rev}.
Moreover, both the Hall coefficient and magnetoresistance
are strongly enlarged due to the spin fluctuation-driven quasiparticle scattering
\cite{Kontani-Hall,Kontani-thermoelectric,Kontani-Nernst-magresi}.
In recent years,
the axial and uniform charge-density-wave (CDW) is observed
in various optimally- and under-doped cuprate superconductors
\cite{CDW_Ghiringhelli,CDW_Chang,CDW_Fujita,uniform_CDW_Matsuda}.
The discovery of CDW has activated the study of the present field.
To explain the CDW mechanism,
spin-fluctuation-driven CDW formation mechanisms have been proposed
\cite{Chubukov_CDW,Kivelson_CDW,Sachdev_CDW,Onari-CDW,Yamakawa-CDW,Kawaguchi-CDW}.

In many previous studies, electronic states in bulk systems with translational symmetry have been analyzed.
On the other hand,
real space structures such as surfaces, interfaces, and impurities
break the translational symmetry of a system,
and they can induce interesting phenomena that cannot be realized in the bulk systems.
In the normal states of cuprate superconductors, YBa$_2$Cu$_3$O$_{7-x}$ (YBCO) and La$_{2-\delta}$Sr$_\delta$CuO$_4$ (LSCO),
non-magnetic impurities induce a local magnetic moment around them,
and the uniform spin susceptibility exhibits the Curie-Weiss behavior \cite{Alloul99-2,Ishida96,Alloul94,Alloul00,Alloul00-2,Alloul99}.
In theoretical studies,
various analyses are performed using the Heisenberg and Hubbard models containing a non-local impurity,
and the enhancement in the spin fluctuations is obtained \cite{Bulut89,Sandvik,Bulut01,Bulut00}.
In case of a local impurity,
the enhancement in the local spin susceptibility is reproduced
by the improved fluctuation-exchange (FLEX) approximation performed in the real space \cite{Kontani-imp}.
Because these analyses are performed in the real space,
the site-dependence of the spin susceptibility is satisfactorily explained.

Recently, the present authors predicted theoretically that
ferromagnetic (FM) fluctuations develop at the open ($1,1$) edge of 	
the two-dimensional cluster Hubbard model.
In addition, as the temperature decreases,
the local mass-enhancement factor
and quasiparticle damping increase strongly at the ($1,1$) edge,
and the system approaches the magnetic critical point. 
The above are edge-induced quantum critical phenomena \cite{Matsubara-edge}.
These impurity or edge-induced magnetic criticalities
originate from the high local density of states (LDOS) sites caused by the Friedel oscillation.
Moreover, the enhanced spin fluctuations may cause
interesting phenomena such as edge-induced spin triplet SC.

On the other hand,
surfaces or interfaces also cause various interesting phenomena in the superconducting (SC) state.
At the ($1,1$) edge or interface of $d_{x^2-y^2}$-wave superconductors,
the Andreev bound state (ABS) is formed, 
and the LDOS increases at the Fermi level
\cite{Hu-ZBCP,Tanaka-ZBCP,Kashiwaya-junction,Matsumoto-Shiba-ABS,Nagato,Kashiwaya-ZBCP}.
This originates from the
sign change in the bulk $d$-wave SC gap.
The ABS is observed by scanning tunneling spectroscopy (STS) experiments as the zero-bias conductance peak
\cite{Kashiwaya-ZBCP-2,Iguchi,Wei-ZBCP,Geek-ZBCP}.
The surface ABS is also regarded
as the odd-frequency pairing amplitude
induced at the surface of an even-frequency superconductor
\cite{Tanaka-odd-frequency,Tanaka-odd-frequency-2}.
Owing to the increase in the LDOS caused by the ABS,
a strong electron correlation is expected to emerge near the edge.
However, theoretical studies on the effects of an ABS on this electron correlation have been limited.
Furthermore,
a surface or an interface can induce a time-reversal symmetry breaking (TRSB) SC state.
For example, it is proposed that the ($1,1$) edge of a $d$-wave superconductor exhibits $d\pm is$-wave SC.
\cite{Matsumoto-Shiba-I,Matsumoto-Shiba-II,Matsumoto-Shiba-III}.
In this case, the relative phase between the $s$- and $d$-wave gaps is $\pi/2$. 
The emergence of TRSB SC state has been discussed
in polycrystalline YBCO \cite{Sigrist-Kuboki-TB}
or twined iron-based superconductor FeSe in the nematic phase \cite{Watashige-FeSe-TB}.
To understand such interesting SC at a surface or an interface,
we have to clarify the effect of the ABS on the spin fluctuations, which can mediate surface-induced SC.

In this study, we investigate the prominent effects of the ABS on the surface electron correlation.
For this purpose,
we construct the two-dimensional cluster Hubbard model with the ($1,1$) edge in the bulk $d$-wave SC state,
and calculate the site-dependent spin susceptibility
by performing random-phase-approximation (RPA) and
modified FLEX approximation ($GV^I$-FLEX) in the real space \cite{Kontani-imp}.
We find that the strong FM fluctuations at the ($1,1$) edge are enhanced much more drastically in
the bulk $d$-wave SC state than in the normal state.
The strong FM fluctuations induced by the surface ABS may drive interesting emerging phenomena, such as edge-induced SC.

%%%%%%%%%%%%%%%%%%%%%%%%%%%%%%%
\section{Model}
\begin{figure}[h]
\includegraphics[width=0.95\linewidth]{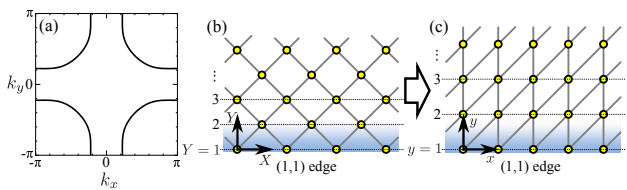}
\caption{(color online)
(a) Fermi surface in the bulk YBCO TB model at filling $n=0.95$.
(b) Square lattice with a ($1,1$) edge.
(c) One-site unit cell square lattice with a ($1,1$) edge.
To simplify the calculation, we use the square lattice shown in (c) instead of that in (b).
Solid lines represent the nearest neighbor bond.
Layer $Y$ in (b) corresponds to layer $y$ in (c).
} 
\label{fig:fig1}
\end{figure}

In this study, we analyze the square-lattice cluster Hubbard model
with a $d$-wave SC gap:
%-Hamiltonian-------------------------------------
\begin{eqnarray}
H&=&\sum_{i,j,\s}t_{i,j}c_{i\s}^\dagger c_{j\s}
+U\sum_{i}n_{i\uparrow}n_{i\downarrow}
\nonumber\\
&&+\sum_{i,j}
\Delta^d_{i,j}
\left(
c_{i\uparrow}^\dagger c_{j\downarrow}^\dagger
+
c_{j\downarrow}c_{i\uparrow}
\right),
\label{eqn:Hamiltonian}
\end{eqnarray}
%-------------------------------------------------
where $t_{i,j}$ is the hopping integral between sites $i$ and $j$.
We set the nearest, next nearest, and third-nearest hopping integrals as $(t,t',t'')=(-1,1/6,-1/5)$,
which correspond to the YBCO tight-binding (TB) model.
$c_{i\sigma}^\dag$ and $c_{i\sigma}$ are the creation and annihilation operators of an electron with spin $\sigma$, respectively.
$U$ is the on-site Coulomb interaction,
and $\Delta^d_{i,j}\equiv\Delta^{d,\uparrow\downarrow}_{i,j}$ is the $d$-wave SC gap between sites $i$ and $j$.
Figure \ref{fig:fig1}(a) shows the Fermi surface of the periodic Hubbard model at filling $n=0.95$.
Then, AFM fluctuations develop owing to the $Q=(\pi,\pi)$ nesting.

In this study, we investigate a
cluster Hubbard model with an open ($1,1$) edge.
Figure \ref{fig:fig1}(b) shows the square lattice with the ($1,1$) edge.
$Y=1$ corresponds to the edge layer.
For convenience, in this study,
we analyze the one-site unit cell structure
shown in Figure \ref{fig:fig1}(c).
This model is periodic along the $x$-direction,
whereas the translational symmetry is violated along the $y$-direction.
%===============================
%===============================
By performing a Fourier transformation on the $x$-direction,
the first term of $(1)$ is expressed as
%-tight-binding-------------------------------------
\begin{eqnarray}
H^0=
\sum_{k_x,y,y',\s}H_{y,y'}^0(k_x)c_{k_x,y,\s}^\dagger c_{k_x,y',\s} .
\label{eqn:tight-binding}
\end{eqnarray}
%------------------------------------------------------------------
We also perform a Fourier transformation on the $x$-direction of the
$d$-wave gap
$
\Delta^d_{i,j}=
\Delta^d/2
(
\delta_{x,x'+1}
\delta_{y,y'+1}
+
\delta_{x,x'-1}
\delta_{y,y'-1}
-
\delta_{x,x'}
\delta_{y,y'+1}
-
\delta_{x,x'}
\delta_{y,y'-1}
)
$.
Here, we assume that
$\Delta^d$ is real and non-zero only between the nearest sites.
Its $(k_x,y,y')$ representation is given as
%-gap Fourier transformation-------------------------------------
\begin{eqnarray}
&&\Delta_{y,y'}^d(k_x,T)
\nonumber\\
&&
=\Delta^d(T)
\left\{
\frac{e^{-ik_x}-1}{2}\delta_{y,y'+1}
+\frac{e^{ik_x}-1}{2}\delta_{y,y'-1}
\right\},
\label{eqn:d-gap}
\end{eqnarray}
%------------------------------------------------------------------
where $\Delta^d(T)$ is the temperature-dependence of the $d$-wave gap function.
We suppose that $\Delta^d(T)$ obeys the BCS-like $T$-dependence:
%-T-dependence of d-wave gap-------------------------------------
\begin{eqnarray}
\Delta^d(T)
=\Delta^d_0
\tanh
\left(
1.74
\sqrt{\frac{T_{cd}}{T}-1}
\right),
\label{eqn:gap_t-dep}
\end{eqnarray}
%------------------------------------------------------------------
where $\Delta^d_0\equiv\Delta^d_0(T=0)$.
Now, we denote the number of sites along $y$-direction as $N_y$.
The $N_y\times N_y$ Green functions
in the $d$-wave SC state, $\hat{G}$, $\hat{F}$ and $\hat{F}^\dag$,
are given as
%-Green function-------------------------------------
\begin{eqnarray}
&&\!\!\!\!\!\!\!\!
\left(
\begin{array}{cc}
\hat{G}(k_x,\e_n) &\hat{F}(k_x,\e_n) \\
\hat{F}^{\dag}(k_x,\e_n) & -\hat{G}(k_x,-\e_n)  
\end{array}
\right)
\nonumber\\
&&\!\!\!\!\!\!\!\!
=
\left(
\begin{array}{cc}
\e_n\hat{1}-\hat{H}^0(k_x)& -{\hat{\Delta}^d}(k_x) \\
     -{\hat{\Delta}^d}(k_x) & \e_n\hat{1}+\hat{H}^0(k_x)
\end{array}
\right)^{-1},
\nonumber \\
\label{eqn:green}
\end{eqnarray}
%------------------------------------------------------------------
where $\e_n=(2n+1)\pi iT$ is the fermion Matsubara frequency.
Here, $(\hat{H}^0)_{y,y'}=H^0_{y,y'}$.

To demonstrate the emergence of the ABS
at the ($1,1$) edge of the TB model in the bulk $d$-wave SC state,
we calculate the LDOS given by
%-LDOSの-------------------------------------
\begin{eqnarray}
\displaystyle D_y(\e)= \frac1{2\pi^2}\int_{-\pi}^{\pi}{d k_x} 
{\rm Im} G_{y,y}(k_x,\e-i\delta).
\label{eqn:ldos}
\end{eqnarray}
%------------------------------------------------
%===============================
% ldos, Andreev bound state
\begin{figure}[h]
\includegraphics[width=0.7\linewidth]{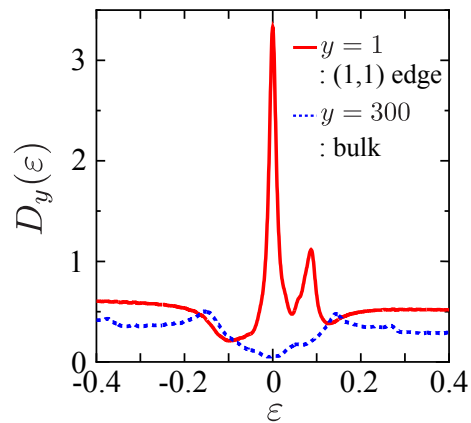}
\caption{(color online)
LDOS in the ($1,1$) edge cluster Hubbard model in the $d$-wave SC state for $\Delta^d=0.08$.
The unit of energy is $|t|=1$.
$y=1$ and $y=300$ correspond to the ($1,1$) edge and bulk, respectively.
For convenience, we set $\delta=0.01$.
}
\label{fig:fig2}
\end{figure}
%===============================
Figure \ref{fig:fig2} displays the obtained LDOS for $\Delta^d(T)=0.08$ by setting $\delta=0.01$.
At the edge ($y=1$), $D_y(\e)$ has a large peak at the Fermi level, $\e=0$, owing to the ABS.
The LDOS at $y=300=N_y/2$ exhibits a V-shape $\e$-dependence,
which corresponds to the bulk LDOS in the $d$-wave SC state.
Note that the height of the peak is proportional to the size of the bulk $d$-wave gap \cite{Matsumoto-Shiba-II}.
A secondary minor peak at $\varepsilon=0.1$ originates from a superconducting surface state
that is different from the surface ABS. We explain the origin of this secondary peak in Appendix A.

%%%%%%%%%%%%%%%%%%%%%%%%%%%%%%%
\section{RPA analysis}
\subsection{formalism}
%RPA
%===============================
\begin{figure}[h]
\includegraphics[width=0.4\linewidth]{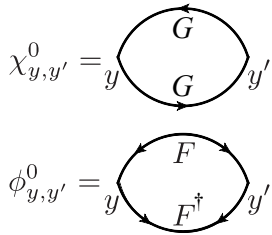}
\caption{
Diagram of the irreducible susceptibility, $\chi^0$ or $\phi^0$,
in the $(k_x,y,y')$ representation.
The line with an arrow is
$G$.
The line with two arrows is $F$ or $F^\dag$.
}
\label{fig:fig3}
\end{figure}
%===============================
In this section,
we calculate the spin susceptibility of the ($1,1$) edge cluster Hubbard model 
using the RPA
in the $(k_x,y,y')$ representation.
Figure \ref{fig:fig3} shows the diagrams of the irreducible susceptibilities, ${\hat \chi}^0$ and ${\hat \phi}^0$.
They are given by ${\hat G}$, ${\hat F}$, and ${\hat F}^\dagger$ as
%--chi^0----------------------------------------------
\begin{eqnarray}
\chi^0_{y,y'}({q}_x,\w_l) &=&-T\sum_{{k}_x,n}
G_{y,y'}({q}_x+{k}_x,\w_l+\e_n)
\nonumber\\
&&\times G_{y',y}({k}_x,\e_n) ,
\label{eqn:chi0}
\end{eqnarray}
%------------------------------------------------
%--phi^0----------------------------------------------
\begin{eqnarray}
\phi^0_{y,y'}(q_x,\omega_l)
&=&
-T
\sum_{k_x,n}
F_{y,y'}(q_x+k_x,\omega_l+\e_n)
\nonumber\\
&&\times
F_{y',y}^{\dag}(k_x,\e_n) ,
\label{eqn:phi0}
\end{eqnarray}
%------------------------------------------------
where  $\omega_l=2l\pi iT$ is the boson Matsubara frequency.
$\phi^0$ is finite only in the SC state.
%RPA
The $N_y\times N_y$ matrix of 
the spin susceptibility ${\hat \chi}$ is calculated
using ${\hat \chi}^0$ and ${\hat \phi}^0$ as
%---Phi---------------------------------------------
\begin{eqnarray}
&&
\hat{\Phi}(q_x,\omega_l)
=
\hat{\chi}^0(q_x,\omega_l)+\hat{\phi}^0(q_x,\omega_l),
\label{eqn:phisc}
\end{eqnarray}
%--------------------------------------------------------
%---RPA chi^s-----------------------------
\begin{eqnarray}
\hat{\chi}(q_x,\omega_l)
&=&
\hat{\Phi}(q_x,\omega_l)
\left\{
\hat{1}-U
\hat{\Phi}(q_x,\omega_l)
\right\}^{-1}.
\label{eqn:chis}
\end{eqnarray}
%--------------------------------------------------------
The spin Stoner factor, $\a_S$, is obtained as the largest eigenvalue
of  $U\hat{\Phi}(q_x,\omega_l)$ at $\w_l=0$.
The magnetic order is realized when $\a_S\ge1$.
Also, the charge susceptibility is
%---RPA chi^c-----------------------------
\begin{eqnarray}
\hat{\chi}^{c}(q_x,\omega_l)
&=&
\hat{\Phi}^{c}(q_x,\omega_l)
\left\{
\hat{1}+U
\hat{\Phi}^{c}(q_x,\omega_l)
\right\}^{-1},
\label{eqn:chic}
\end{eqnarray}
%--------------------------------------------------------
where $\hat{\Phi}^{c}(q_x,\omega_l)=\hat{\chi}^0(q_x,\omega_l)-\hat{\phi}^0(q_x,\omega_l)$.

\subsection{Numerical result of $\hat{\chi}$ and $\alpha_S$ in real space}
Next, we perform the RPA
analyses for the 
cluster Hubbard model with the bulk $d$-wave SC gap, with the translational symmetry along the $x$-direction.
We set the number of $k_x$-meshes as $N_x=64$,
that of sites along the $y$-direction as $N_y=64$
and that of Matsubara frequencies as $N_\omega=1024$.
We set the electron filling, $n=0.95$; the Coulomb interaction is $U=2.25$ in the RPA.
Here, the unit of energy is $|t|=1$, which corresponds to
$\sim 0.4$eV in cuprate superconductors.
We set the transition temperature for the $d$-wave SC as $T_{cd}=0.04$.
In addition, we define $\Delta_{\rm max}$ as the maximum value of the $d$-wave gap on the Fermi surface.
In the present model, $\Delta_{\rm max}=1.76\Delta_0^d$ for $n=0.95$.
Experimentally, $4<2\Delta_{\rm max}/T_{cd} <10$ in YBCO
\cite{cuprate_coherence_1,cuprate_coherence_2}.
Therefore, we set $\Delta_0^d=0.06$ or 0.09, which corresponds to $2\Delta_{\rm max}/T_{cd}=5.28$ or 7.92 for $T_{cd}=0.04$.
By performing this analysis,
we show that the ABS drastically enhances the FM fluctuations at the ($1,1$) edge,
and the system rapidly approaches a magnetic-order phase.

%****RPA************************************************************************************
First, we study the site-dependent static spin susceptibility, $\hat{\chi}(q_x,\w_l=0)$, 
in the $d$-wave SC state using the RPA.
Hereafter, we refer to the spin susceptibility in the normal state as $\hat{\chi}^{(n)}$.
We also introduce the following susceptibilities in the SC state
to clarify the origin of the enhancement in the FM fluctuations:
%-\chi''---------------------------------------------------------------
\begin{eqnarray}
{\hat \chi'}={\hat \Phi}'(1-U{\hat \Phi}')^{-1} \quad ({\hat \Phi}'={\hat \chi}^0),
\label{eqn:chi'}
\\
{\hat \chi}''={\hat \Phi}''(1-U{\hat \Phi}'')^{-1} \quad ({\hat \Phi}''={\hat \chi}^{0(n)}+{\hat \phi}^0).
\label{eqn:chi''}
\end{eqnarray}
%----------------------------------------------------------------
Here, ${\hat \chi^0}$ and ${\hat \chi}^{0(n)}$
are the irreducible susceptibilities in the bulk $d$-wave SC and normal states, respectively.
In susceptibilities ${\hat \chi'}$ and ${\hat \chi''}$,
the effect of $d$-wave gap in ${\hat \phi}^0$ and ${\hat \chi}^0$ are dropped, respectively.

\begin{figure}[h]
\includegraphics[width=0.7\linewidth]{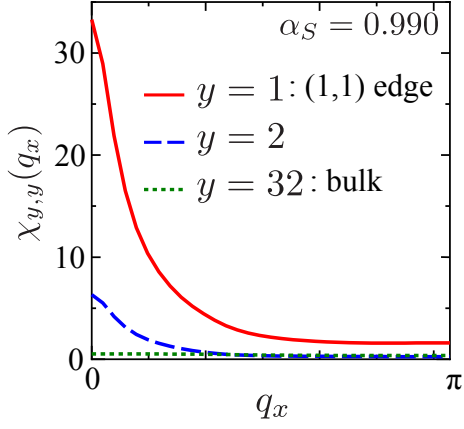}
\caption{(color online)
$q_x$-dependence of $\chi_{y,y}(q_x)$ obtained by the RPA for
$\Delta_0^d=0.09$ at $T=0.0365$.
The value of $y$ corresponds to the depth from the ($1,1$) edge.
$y=1$ is the ($1,1$) edge and $y=32$ corresponds to the bulk.
}
\label{fig:fig4}
\end{figure}

% site-dependence of chi
Figure \ref{fig:fig4} shows the obtained RPA susceptibilities for
$\Delta_0^d=0.09$ at $T=0.0365$.
$\chi_{y,y}(q_x)$ represents the correlation of the spins in the same layer $y$ at $\omega_l=0$.
In the edge layer ($y=1$), $\chi_{y,y}(q_x)$ has a large peak at $q_x=0$.
This result means that strong FM fluctuations develop in the ($1,1$) edge layer.
The FM correlation along the edge layer is consistent with the AFM correlation in the periodic Hubbard model.
This strong enhancement occurs only for $y=1$ and $y=2$.
In fact, the Stoner factor is $\alpha_S=0.990$ with the edge, whereas $\a_S=0.673$
in the periodic model.
Therefore, the present model with the bulk $d$-wave SC gap approaches the magnetic quantum critical point
with introduction of the edge.
%===============================
% site-dependence spin susceptibility of RPA 
%===============================

\begin{figure}[h]
\includegraphics[width=0.7\linewidth]{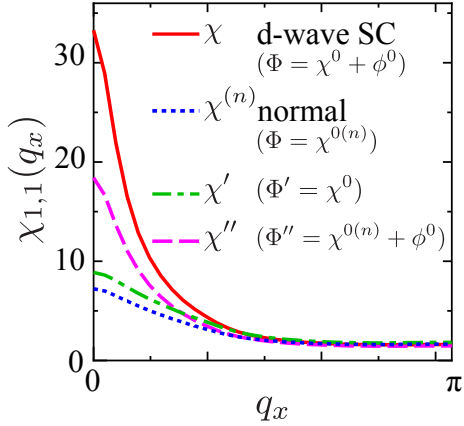}
\caption{(color online)
Comparison between $\chi_{1,1}(q_x)$, $\chi^{(n)}_{1,1}(q_x)$, $\chi'_{1,1}(q_x)$, and $\chi''_{1,1}(q_x)$
for $\Delta_0^d=0.09$ at $T=0.0365$.
}
\label{fig:fig5}
\end{figure}

%normal
Next, we compare the $d$-wave SC and normal state.
Figure \ref{fig:fig5} shows $\chi_{1,1}$ and $\chi_{1,1}^{(n)}$ in the model with edge.
The enhancement in the FM fluctuations is much more drastic in the $d$-wave SC state
compared to that in the normal state discussed in Ref. \cite{Matsubara-edge}.
Therefore, this strong enhancement cannot be explained only by the existence of edge.

% chi', chi''
Furthermore, we examine the contribution from ${\hat \phi}^{0}$ and ${\hat \chi}^0$
to the enhancement of total spin susceptibility.
In Figure \ref{fig:fig5}, we present $\chi'_{1,1}(q_x)$ and $\chi''_{1,1}(q_x)$.
The height of the peak of ${\hat \chi}'$ is much smaller than that of ${\hat \chi}$.
On the other hand, the height of the peak of ${\hat \chi}''$ is enlarged whereas slightly lower than that of ${\hat \chi}$.
Therefore, ${\hat \phi}^0$ due to anomalous Green functions gives the dominant contribution for the increment of $\hat{\chi}$.
Also $\hat{\chi}^0-\hat{\chi}^{0(n)}$ gives minor contribution since ${\hat \chi}'>\hat{\chi}^{(n)}$.
%===============================
% normal chi', chi''
%===============================

\begin{figure}[h]
\includegraphics[width=0.95\linewidth]{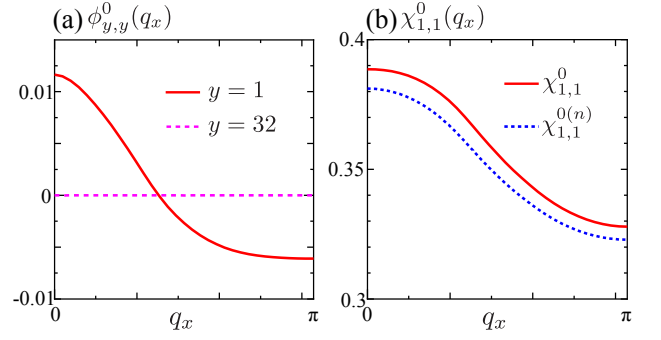}
\caption{(color online)
$q_x$-dependence of irreducible susceptibilities
for $\Delta_0^d=0.09$ at $T=0.0365$.
(a) Comparison between $\phi_{y,y}^0(q_x)$ at the edge ($y=1$) and in the bulk ($y=32)$.
(b) Comparison between $\chi_{1,1}^0(q_x)$ and $\chi_{1,1}^{0(n)}(q_x)$.
}
\label{fig:fig6}
\end{figure}

% irreducible susceptibility
Figure \ref{fig:fig6}(a) shows the $q_x$-dependence of
${\hat \phi}^0$.
In the bulk, $\phi^{0}_{32,32}$ is zero
because $x$-axis is the direction of the $d$-wave gap node.
Interestingly, $\phi_{1,1}^0$ is finite and has a peak at $q_x=0$.
This is explained as an effect of the ABS, which corresponds to
the odd-frequency SC induced at the ($1,1$) edge
as discussed in Refs. \cite{Tanaka-odd-frequency,Tanaka-odd-frequency-2}.
We give brief discussion on this issue in Appendix B.
In Figure \ref{fig:fig6}(b),
we show the $q_x$-dependence of
${\hat \chi}^{0}$ and ${\hat \chi}^{0(n)}$.
At the edge, $\chi^{0}_{1,1}$ is slightly larger than $\chi^{0(n)}_{1,1}$ owing to the peak of LDOS due to the ABS.
%===============================
% irreducible susceptibility of RPA
%===============================

\begin{figure}[h]
\includegraphics[width=0.8\linewidth]{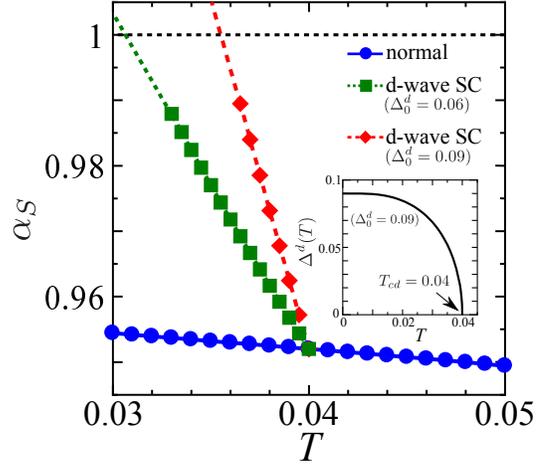}
\caption{(color online)
$T$-dependence of $\alpha_S$ in the RPA.
The inset shows the $T$-dependence of the size of the $d$-wave gap.
We assume that the BCS-like $T$-dependence given by \eqref{eqn:gap_t-dep}.
We set the transition temperature of the $d$-wave SC as $T_{cd}=0.04$.
}
\label{fig:fig7}
\end{figure}

%T-dependence in RPA
Figure \ref{fig:fig7} shows the $T$-dependence of $\a_S$ in the RPA.
The inset shows the $T$-dependence of the size of the $d$-wave gap, which is given in Eq.\eqref{eqn:gap_t-dep}.
$\a_S$ in the SC state increases sharply as $T$ decreases compared to that in the normal state,
due to the development of the ABS.
The increase for $\Delta_0^d=0.09$ is sharper than that for $\Delta_0^d=0.06$
because the height of the ABS is proportional to $\Delta_0^d$.
$\a_S$ reaches unity at $T\approx0.036$ for $\Delta_0^d=0.09$,
and the edge FM order is realized.
To summarize, we predict the emergence of FM order at ($1,1$) edge of $d_{x^2-y^2}$-wave superconductors.
%===============================
% T-dependence of alpha_s of RPA
%===============================

%  FLEX
%%%%%%%%%%%%%%%%%%%%%%%%%%%%%%%
\section{FLEX analysis}
\subsection{$GV^I$-FLEX}
In this section, we study the spin susceptibility using modified FLEX ($GV^I$-FLEX) approximation
developed in Ref. \cite{Kontani-imp},
since the negative feedback effect on $\hat{\chi}$ near the impurity is prominently overestimated in conventional FLEX.
In fact, the negative feedback is suppressed
by vertex corrections near the impurity as pointed out in Ref. \cite{Kontani-imp}.
In the modified FLEX,
the cancellation between negative feedback and vertex corrections
is assumed, and then reliable results are obtained for single impurity problem \cite{Kontani-imp}.

% GVI-FLEX
To apply the modified FLEX to the present model,
we first calculate the self-energy in the periodic system without the edge, $\Sigma^0(k_x,k_y,i\varepsilon_n)$,
using the conventional FLEX approximation.
Then, by performing the Fourier transformation for $y$-direction,
we obtain $\Sigma^0(k_x,y,y',i\varepsilon_n)=\Sigma^0(k_x,y-y',i\varepsilon_n)$.
Next, we calculate the Green functions in the ($1,1$) edge model with $\Sigma^0(k_x,y,y',i\varepsilon_n)$
%-GVI Green functions-------------------------------------
\begin{eqnarray}
&&\!\!\!\!\!\!\!\!
\left(
\begin{array}{cc}
\hat{G^I}(k_x,\e_n) &\hat{F^I}(k_x,\e_n) \\
\hat{F^I}^{\dag}(k_x,\e_n) & -\hat{G^I}(k_x,-\e_n)  
\end{array}
\right)
\nonumber\\
&&\!\!\!\!\!\!\!\!
=
\left(
\begin{array}{cc}
\e_n\hat{1}-\hat{H}^0(k_x)-\hat{\Sigma}^0(k_x,i\varepsilon_n)& -{\hat{\Delta}^d}(k_x) \\
     -{\hat{\Delta}^d}(k_x) &\!\!\!\!\!\!\!\!\!\!\!\!\!\!\!\!\!\!\!\!\!\! \e_n\hat{1}+\hat{H}^0(k_x)+\hat{\Sigma}^0(k_x,-i\varepsilon_n)
\end{array}
\right)^{-1},
\nonumber \\
\label{eqn:green}
\end{eqnarray}
%------------------------------------------------------------------
where $\hat{H}^0(k_x)$ is the tight-binding model with the ($1,1$) edge.
In the $GV^I$-FLEX, the spin susceptibility is calculated by
$\hat{G^I}$, $\hat{F^I}$, and $\hat{F^I}^\dag$
instead of $\hat{G}$, $\hat{F}$, and $\hat{F}^\dag$ in Eqs.\eqref{eqn:chi0}--\eqref{eqn:chis}.

% renormalization
In this approximation,
the $d$-wave gap is renormalized by the self-energy.
The renormalized $d$-wave gap in bulk
is evaluated by ${\Delta_0^d}^*=\Delta_0^d/Z_{\rm{bulk}}$,
where $Z_{\rm{bulk}}$ is the on-site mass-enhancement factor in the bulk.

% GVI-FLEX
\subsection{Numerical result of $\hat{\chi}$ and $\alpha_S$ in real space}
%GVI-FLEX parameter
In the numerical study of $GV^I$-FLEX, we set the number of $k_x$-meshes as $N_x=64$,
that of sites along the $y$-direction as $N_y=64$,
and that of Matsubara frequencies as $N_\omega=1024$.
We set the electron filling, $n=0.95$; the transition temperature for the $d$-wave is $T_{cd}=0.04$.
The Coulomb interaction is $U=2.65$.

\begin{figure}[h]
\includegraphics[width=0.7\linewidth]{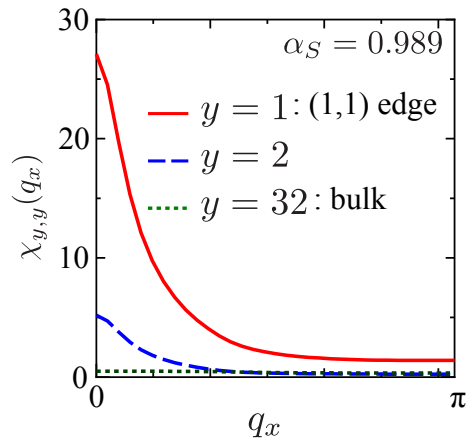}
\caption{(color online)
$q_x$-dependence of $\chi_{y,y}(q_x)$ obtained
by the $GV^I$-FLEX for $\Delta_0^d=0.12$ at $T=0.036$.
$y=32$ corresponds to the bulk.
With this parameter, we obtain $Z_{\rm bulk}=1.37$.
The renormalized gap is ${\Delta_0^d}^*\approx0.087$
and $2{\Delta_{\rm max}}^*/T_{cd}\approx7.69$.
}
\label{fig:fig8}
\end{figure}

Figure \ref{fig:fig8} shows the $q_x$-dependence of $\chi_{y,y}(k_x)$
in the $GV^I$-FLEX for $\Delta_0^d=0.12$ at $T=0.036$.
With this parameter,
we obtain $Z_{\rm bulk}=1.37$, ${\Delta_0^d}^*\approx0.087$ and $2{\Delta_{\rm max}}^*/T_{cd}\approx7.69$.
At the ($1,1$) edge ($y=1$), $\chi_{1,1}(q_x)$ has a large peak at $q_x=0$.
In the periodic model without edge, $\alpha_S=0.699$ in the FLEX approximation.
The Stoner factor increases to $\alpha_S=0.989$ by introducing the ($1,1$) edge.
Therefore, the enhancement in the FM fluctuations at the edge is obtained
by both the RPA and $GV^I$-FLEX.
%===============================
% site-dependence of xs of FLEX
%===============================

\begin{figure}[h]
\includegraphics[width=0.8\linewidth]{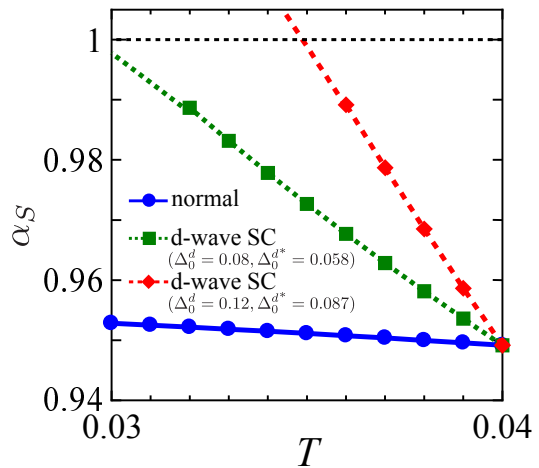}
\caption{(color online)
$T$-dependence of $\alpha_S$
in the $GV^I$-FLEX.
We set the transition temperature of the $d$-wave SC as $T_{cd}=0.04$.
We obtained
${\Delta_0^d}^*\approx0.058$ and $2{\Delta_{\rm max}}^*/T_{cd}\approx5.11$ for $\Delta_0^d=0.08$.
}
\label{fig:fig9}
\end{figure}

%T-dependence in GVI-FLEX
Figure \ref{fig:fig9} shows the $T$-dependence of $\alpha_S$
in ($1,1$) edge cluster model given by the $GV^I$-FLEX.
In the normal state, $\alpha_S$ increases gently as $T$ decreases.
However, in the presence of bulk $d$-wave SC gap,
$\alpha_S$ increases sharply as $T$ decreases.
For $\Delta_0^d=0.08$, the mass-enhancement factor is $Z_{\rm bulk}=1.38$ at $T=0.032$.
Thus, we obtain ${\Delta_0^d}^*\approx0.058$ and $2{\Delta_{\rm max}}^*/T_{cd}\approx5.11$.
For $\Delta_0^d=0.12$, $\alpha_S$ reaches 0.99 at $T=0.036$.
For a fixed ratio
$2{\Delta_{\rm max}}^*/T_{cd}$,
the obtained $T$-dependence of $\alpha_S$ is comparable in both the RPA and $GV^I$-FLEX.
%===============================
% T-dependence of alpha_s of FLEX
%===============================

%***********************************************************
\section{Effect of the finite $d$-wave coherence length on the edge-induced spin fluctuations}
\begin{figure}[h]
\includegraphics[width=0.95\linewidth]{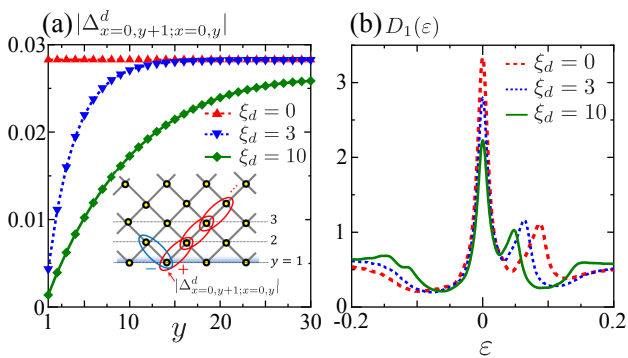}
\caption{(color online)
(a) Site-dependence of the $d$-wave gap suppressed near the edge over $\xi_d$.
We set $\Delta_{0}^{d}=0.08$, and plot it at $T=0.032$.
$\xi_d=0$ corresponds to the site-independent $d$-wave gap.
The inset shows the nearest neighbor bonds corresponding to $|\Delta^d_{x=0,y+1;x=0,y}|$.
(b) LDOS at the ($1,1$) edge for the finite $\xi_d$.
}
\label{fig:fig10}
\end{figure}

In this section, we study the enhancement in the FM fluctuations 
when the $d$-wave gap is suppressed near the edge for a finite range, $1\leq y\leq \xi_d$,
where $\xi_d$ is the coherence length of the $d$-wave SC.
We set the $y$-dependence of the $d$-wave gap as 
%-------------------------------------------------------------------
\begin{eqnarray}
\Delta_{y,y'}^d(k_x,T)\left\{1-\exp\left(\frac{y+y'-2}{2\xi_d}\right)\right\}.
\label{eqn:gap_xd}
\end{eqnarray}
%--------------------------------------------------------------------
We note that,
the anomalous self-energy for the $d$-wave SC gap
is calculated self-consistently in the SC FLEX approximation below $T_{cd}$ \cite{Takimoto_FLEX}.
If the SC FLEX is applied to the edge cluster model,
the $d$-wave gap for $y\lesssim \xi_d$ should be naturally suppressed.
Instead, we set $\xi_d$ as a parameter to simplify the analysis.
From the experimental results
\cite{cuprate_coherence_3,cuprate_coherence_4,cuprate_coherence_5,cuprate_lattice_1},
we can estimate $\xi_d$ to be 3 sites for $T\ll T_{cd}$.
For $T\lesssim T_{cd}$, 
$\xi_d\gg 3$ because of the relation $\xi_d \propto (1-T/T_{cd} )^{-1/2}$ in the GL theory.
Thus, we set $\xi_d=3$ and 10 in this analysis.
The $y$-dependence of
given $|\Delta^d_{x=0,y+1;x=0,y}|$
is shown in Figure \ref{fig:fig10}(a).
The inset shows the corresponding nearest neighbor bonds in the real space.
Figure \ref{fig:fig10}(b) shows the LDOS at the edge.
Although the height of the peak of the ABS is reduced,
the peak structure remains for finite $\xi_d~(\lesssim 10)$.
%===============================
% finite xi_d
%===============================

\begin{figure}[h]
\includegraphics[width=0.95\linewidth]{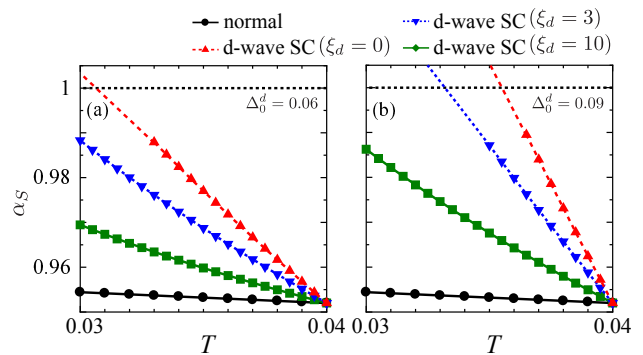}
\caption{(color online)
$T$-dependence of $\alpha_S$ by the RPA
for (a) $\Delta_0^d=0.06$ or (b) 0.09
with finite $\xi_d$.
The red dashed line represents $\alpha_S$ for the site-independent $d$-wave gap.
The black solid line represents $\alpha_S$ in the normal state.
}
\label{fig:fig11}
\end{figure}

% alpha_S
Next, we calculate the $T$-dependence of $\alpha_S$ using the RPA,
and Figure \ref{fig:fig11} shows the result
for (a) $\Delta_0^d=0.06$ and (b) 0.09 for $\xi_d=0$--$10$.
The increase in $\alpha_S$ for $\xi_d=10$ is moderate compared to that for $\xi_d=0$ and 3,
owing to the suppression of the ABS.
For $\Delta_0^d=0.09$,
$\alpha_S$ reaches 0.986 at $T=0.03$ even for $\xi_d=10$.
However, for $\Delta_0^d=0.06$ and $\xi_d=10$,
the increase in $\alpha_S$ becomes milder
and $\alpha_S\approx0.97$ even at $T=0.03$.
Therefore, we conclude that the drastic enhancement in the FM fluctuations
is realized under the conditions $2\Delta_{\rm max}/T_{cd}\gtrsim 6$ and $\xi_d \ll 10$,
both of which are satisfied in real cuprate superconductors.
%==============================
% T-dependece of alpha_S
%===============================

%%%%%%%%%%%%%%%%%%%%%%%%%%%%%%%
\section{Summary}
In this study,
we revealed that the ABS drastically enhances the FM fluctuations at the ($1,1$) edge of the $d$-wave superconductor.
For this purpose,
we construct the two-dimensional square lattice Hubbard model with the edge in the presence of the bulk $d$-wave SC gap.
Then we perform the site-dependent RPA calculation in the real space.
By detailed analysis,
we found that 
edge-induced FM fluctuations are mainly caused by the increment of $\hat{\phi}^0$ due to the ABS.
Furthermore,
the Stoner factor $\alpha_S$ exhibits drastic increase
just below the bulk $d$-wave $T_{c}$, and edge-induced FM order or fluctuations is expected to emerge.
Such ABS-induced magnetic critical phenomena are also obtained
by using the $GV^I$-FLEX.
Finally, we verified that the the enhancement in FM fluctuations are still prominent
under the conditions $2\Delta_{\rm max}/T_{cd}\gtrsim 6$ and $\xi_d \ll 10$,
which are satisfied in cuprate superconductors.
Therefore, we conclude that the ABS-induced FM order or strong FM fluctuations appears in
real cuprate  superconductors.

\begin{figure}[h]
\includegraphics[width=0.7\linewidth]{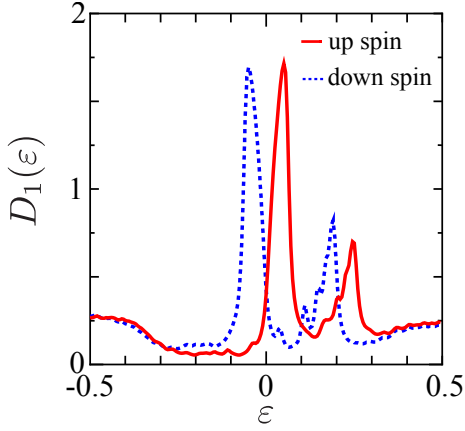}
\caption{(color online)
LDOS at the edge of $d$-wave superconductor ($\Delta^d=0.20$) when the magnetization ($M_0=0.10$) emerges.
The red solid line and blue doted line represent the LDOS for up and down spins, respectively. 
For convenience, we set $\delta=0.01$.
}
\label{fig:fig12}
\end{figure}

The result of the present study indicates the emergence of interesting edge-induced phenomena.
For example, the edge FM order will induce the splitting of the ABS peak, which may be observed by STM/STS study.
Figure \ref{fig:fig12} shows the LDOS for up and down spins at the edge with the magnetization ($M_0=0.10)$.
The magnetization is given by the Zeeman term
%---------------------------------------------------------
$
H_M=
{M_0}/{2}
\sum_{k_x,\sigma}
\sigma
c_{k_x 1 \sigma}^\dag c_{k_x 1 \sigma}.
$
%---------------------------------------------------------
In addition, an edge-induced triplet SC is expected to be realized theoretically \cite{dpip-sc}.
In this case, the bulk $d$-wave SC and edge-induced triplet SC may coexist at the ($1,1$) edge ($d\pm ip$-wave),
similar to the $d\pm is$-wave state discussed in Ref. \cite{Matsumoto-Shiba-I,Matsumoto-Shiba-II,Matsumoto-Shiba-III}.
This presents an important problem for the future,
to understand the edge-induced SC state in strongly correlated electron systems.
%==============================
% LDOS magnetic order at the (1,1) edge
%===============================

%%%%%%%%%%%%%%%%%%%%%%%%%%%%%%%
\begin{acknowledgements}
We are grateful to Y. Tanaka, S. Onari, and Y. Yamakawa for valuable comments and discussions.
This work was supported by the JSPS KAKENHI (No. JP19H05825 and No. JP18H01175).
\end{acknowledgements}

%%%%%%%%%%%%%%%%%%%%%%%%%%%%%%%
\appendix
\section{The origin of the minor peak of the LDOS}
\begin{figure}[h]
\includegraphics[width=0.95\linewidth]{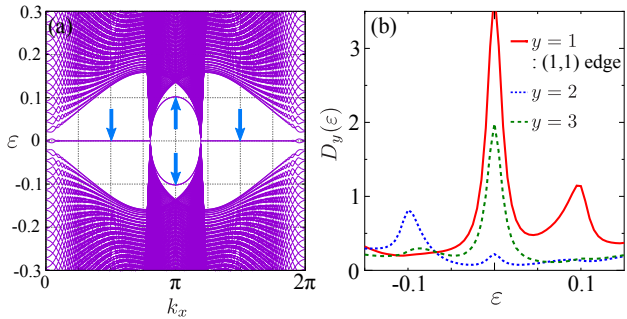}
\caption{(color online)
(a) Energy spectra of the $d$-wave SC cluster model with the ($1,1$) edge.
The flat dispersion at $\varepsilon=0$ is the ABS.
There are two surface states separated from the bulk states
in the range of $3\pi/4\lesssim k_x \lesssim5\pi/4$.
These surface states are pointed out by arrows.
(b) LDOS near the ($1,1$) edge in the bulk $d$-wave SC state.
The minor peaks at $\varepsilon=\pm 0.1$
correspond to the edge states in (a).
}
\label{fig:fig13}
\end{figure}

In this appendix, we explain the origin of the secondary minor peak of the edge LDOS at $\varepsilon=0.1$
shown in Fig. \ref{fig:fig2}.
For this purpose, we calculate the energy spectra of the $d$-wave SC cluster model with the ($1,1$) edge.
Figure \ref{fig:fig13} (a) shows
the obtained energy spectra for $\Delta^d=0.09$ ($\Delta_{\rm max}=0.158$).
The flat dispersion at $\varepsilon=0$ corresponds to the surface ABS.
In addition, there are two surface states separated from the bulk states
in the range of $3\pi/4\lesssim k_x \lesssim5\pi/4$.
These surface states can give minor peak in the LDOS.
As shown in Figure \ref{fig:fig13} (b),
the LDOS at $y=1$ ($y=2$) possesses a minor peak at $\varepsilon=0.1$ ($\varepsilon=-0.1$).
Thus, it is verified that minor peaks at $\varepsilon=\pm 0.1$
in the LDOS originate from the finite-energy surface state in Fig. \ref{fig:fig13} (a).
%===============================
% minor peak
%===============================

\section{Relation between enhanced FM fluctuations and odd-frequency superconductivity}
\begin{figure}[h]
\includegraphics[width=0.95\linewidth]{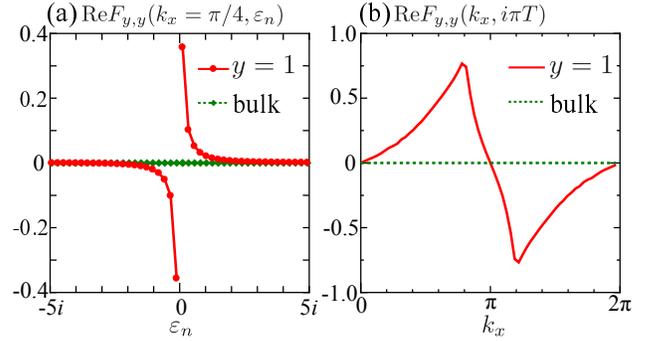}
\caption{(color online)
Anomalous Green function
calculated for $\Delta_{0}^{d}=0.09$ at $T=0.0365$.
(a) $\varepsilon_n$-dependence of ${\rm{Re}}F_{y,y}(k_x=\pi/4,\varepsilon_n)$.
The red and green points represent the component in the edge ($y=1$) and bulk (periodic system), respectively.
(b) $k_x$-dependence of ${\rm{Re}}F_{y,y}(k_x,i\pi T)$.
The red solid line and green doted line represents the component in the edge and bulk, respectively.
}
\label{fig:fig14}
\end{figure}

Here, we discuss the reason for the enhancement of ${\hat \phi}^0$ near the ($1,1$) edge in more detail.
First, we examine
the anomalous Green function,
by which
${\hat \phi}^0$ is composed.
Figure \ref{fig:fig14} (a)
shows the $\varepsilon_n$-dependence of ${\rm{Re}}F_{y,y}(\pi/4,\varepsilon_n)$.
In the bulk, ${\rm{Re}}F_{y,y}(\pi/4,\varepsilon_n)=0$
because the $x$-direction is the node direction of $d$-wave gap.
However, at the edge, ${\rm{Re}}F_{1,1}(\pi/4,\varepsilon_n)$ is finite, and it shows an odd-frequency-dependence.
This odd-frequency pair amplitude
can be understood as another physical picture of the ABS \cite{Tanaka-odd-frequency,Tanaka-odd-frequency-2}.
Figure \ref{fig:fig14} (b) shows the $k_x$-dependence of ${\rm{Re}}F_{y,y}(k_x,i\pi T)$.
At the edge, ${\rm{Re}}F_{1,1}(k_x,i\pi T)$ is finite and has peaks at $k_x\approx 4\pi/5$ and $k_x\approx 6\pi/5$,
whereas ${\rm{Re}}F_{y,y}(k_x,i\pi T)=0$ in the bulk. 
These peaks generate the enhancement of $\phi_{1,1}^0$ at $q_x=0$.
Therefore, the enhancement in the FM fluctuations by ${\hat \phi}^0$
can be explained as the direct effect of the odd-frequency pairing, which is an aspect of the ABS.
%===============================
% kx and matsubara frequency dependence of F
%===============================
%===============================
% odd-frequency
\begin{figure}[h]
\includegraphics[width=0.95\linewidth]{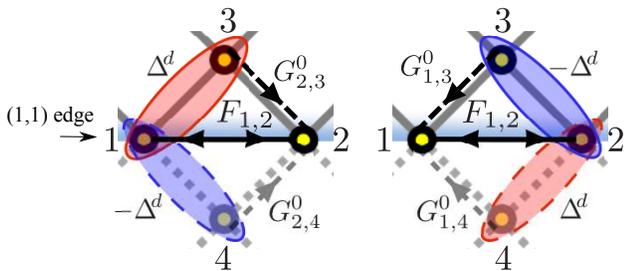}
\caption{(color online)
Contributions to the anomalous Green function at the ($1,1$) edge, $F_{1,2}$.
The solid line with two arrows represents $F_{1,2}$.
The red and blue circle show the $d$-wave SC gap between the nearest neighbor sites.
The doted line with an arrow is the Green function in the normal state, $G^0$.
At the edge, $F_{1,2}$ is finite because the contributions through site 4 are dropped.
}
\label{fig:fig15}
\end{figure}
%===============================

%odd-frequency reason
Next, we discuss why large odd-frequency component appears at the edge based on the atomic picture,
assuming that $t$ is the small parameter.
The zeroth-order Green function at the same site $i$ in the normal state is
%---------------------------------------------------------
\begin{align}
G_{i,i}^0(\varepsilon_n)
=
\frac{1}{\varepsilon_n-E},
\end{align}
%---------------------------------------------------------
where $E$ is the atomic level.
The Green function between the nearest neighbor sites $i$ and $j$
is represented by the first-order perturbation of hopping integral $t$ as follows:
%---------------------------------------------------------
\begin{align}
G_{i,j}^0(\varepsilon_n)
&=
\frac{1}{\varepsilon_n-E}
t
\frac{1}{\varepsilon_n-E}
\nonumber\\
&=
\frac{t}{(\varepsilon_n-E)^2}.
\end{align}
%---------------------------------------------------------
Figure \ref{fig:fig15} shows the lowest order contributions to the anomalous Green function at the edge, $F_{1,2}$.
They are represented as follows:
%---------------------------------------------------------
\begin{align}
F_{1,2}(\varepsilon_n)
=&
-
G_{1,1}^0(\varepsilon_n)
\Delta_{1,3}^d
G_{2,3}^0(-\varepsilon_n)
\nonumber\\
&
-
G_{1,3}^0(\varepsilon_n)
\Delta_{3,2}^d
G_{2,2}^0(-\varepsilon_n)
\nonumber\\
=&
-
\frac{2\Delta_{1,3}^d t \varepsilon_n}{(E^2-\varepsilon_n^2)^2}.
\end{align}
%---------------------------------------------------------
In the second equal sign, we use $\Delta_{1,3}^d=-\Delta_{3,2}^d$.
Therefore, $F_{1,2}(\varepsilon_n)$ is odd for $\varepsilon_n$.
In the bulk, $F_{1,2}$ vanishes
because the contributions through site 4 cancel those through site 3.

%%%%%%%%%%%%%%%%%%%%%%%%%%%%%%%%%%%%%%%%%%%%%%%%%% end

%%%%%%%%%%%%%%%%%%%%
% references
%%%%%%%%%%%%%%%%%%%%
%\newpage

\end{document}